\documentstyle[12pt]{article}
\begin{document}

\begin{center}
\vspace{2cm} {\bf SYNCHRONIZATION IN THE IDENTICALLY DRIVEN SYSTEMS }%
\\[\baselineskip]

{\ }{\small B. Kaulakys}$^1${\small , F. Ivanauskas}$^2${\small \ and T.
Me\v skauskas}$^2${\small \\[\baselineskip] \vspace{-0.5\baselineskip} }$^1$%
{\small Institute of Theoretical Physics and Astronomy, A. Go\v stauto 12,
2600 Vilnius, Lithuania \\}$^2$%
{\small Department of Mathematics, Vilnius University, Naugarduko 24, 2006
Vilnius, Lithuania}\\[\baselineskip]
\end{center}

\begin{quote}
{\it Abstract}: We investigate a transition from chaotic to nonchaotic
behavior and synchronization in an ensemble of systems driven by identical
random forces. We analyze the synchronization phenomenon in the ensemble of
particles mo\-ving with friction in the time-dependent potential and driven by
the identical noise. We define the threshold values of the parameters for
transition from chaotic to nonchaotic behavior and investigate dependencies
of the Lyapunov exponents and power spectral density on the nonlinearity of
the systems and character of the driven force.
\end{quote}

\baselineskip=1.5\normalbaselineskip

\noindent {\bf 1. Introduction}

Trajectories of the nonlinear dynamical systems are very sensitive to
initial conditions and unpredictable. The systems exhibit an apparent
random behavior. It might be expected that turning on an additional random
forces make their behavior even ``more random''. However, as it was shown by
Fahy \& Hamann [1992] and Kaulakys \& Vektaris [1995a,b] when an ensemble
of bounded in a fixed external potential particles with different
initial conditions are driven by an identical sequence of random forces,
the ensemble of trajectories may become identical at long times.
The system becomes not chaotic: the trajectories are independent on the
initial conditions. Here we analyze the similar phenomenon in the ensemble
of particles moving with friction in the time-dependent potential and
driven by the identical noise. We define the threshold values of the
parameters for transition from chaotic to nonchaotic behavior and
investigate dependencies of the Lyapunov exponents and power spectral
density on the nonlinearity of the systems and character of the driven force.

\noindent {\bf 2. Models and Results}

Consider a system of particles of mass $m=1$ moving according to Newton's
equations in the time dependent potential $V({\bf r},t)$, e.g. in the
potential $V\left(x,t\right)=x^4-x^2-ax\sin\omega t$, and with the
friction coefficient $\gamma$. At time intervals $\tau$ the particles are
partially stopped and their velocities are reset to the mixture of
some part $\alpha$ of the old velocities with random velocity
${\bf v}_i^{ran}$: ${\bf v}^{new}=\alpha {\bf v}^{old}+{\bf v}_i^{ran}$,
where $i$ is the stop number. Note that ${\bf v}_i^{ran}$ depends on the
stop number $i$ but not on the particle. The simplest and most natural
way is to choose the random values of velocity ${\bf v}_i^{ran}$ from
a Maxwell distribution with $k_BT=m=1$.

A transition from chaotic to nonchaotic behavior in such a system may be
detected from analysis of the neighboring trajectories of two particles
initially at points ${\bf r}_0$ and ${\bf r}_0^{\prime}$ with starting
velocities ${\bf v}_0$ and ${\bf v}_0^{\prime }$. The convergence of
the two trajectories to the single final trajectory depends on the evolution
with a time of the small variances
$\Delta {\bf r}_i={\bf r}_i^{\prime}-{\bf r}_i$
and $\Delta {\bf v}_i={\bf v}_i^{\prime }-{\bf v}_i$. From formal solutions
${\bf r}={\bf r}({\bf r}_i,{\bf v}_i,t)$ and
${\bf v}={\bf v}({\bf r}_i,{\bf v}_i,t)$ of the Newton's equations with
initial conditions ${\bf r}={\bf r}_i$ and ${\bf v}={\bf v}_i$ at $t=0$
it follows the mapping form of the equations of motion for $\Delta {\bf r}$
and $\Delta {\bf v.}$ The analysis of dynamics based on these equations
has been investigated by Kaulakys \& Vektaris [1995a,b].

Here we calculate the Lyapunov exponents directly from the equations of
motion and linearized equation for the variances and extend the
investigation for the systems with friction in the regular external field
and perturbed by the identical for all particles random force. In Fig.~1 we
show the dependence on $\tau$ of the Lyapunov exponents for the motion in
the nonautonomous Duffing potential with friction described by the equations
$$
   \dot v = 2x - 4x^3 - \gamma v + a\sin\omega t, \qquad \dot x = v. \eqno(1)
$$

For the values of parameters corresponding to the positive Lyapunov
exponents, i.e. without the random perturbation ($\tau \rightarrow \infty $)
the system is chaotic. The negative Lyapunov exponents for small $\tau$
indicate to the nonchaotic Brownian-type motion.

As it was already been observed in [Kaulakys \& Vektaris,~1995b] such
systems exhibit the intermittency route to chaos which provides sufficiently
universal mechanism for $1/f$--type noise in the nonlinear systems.
Here we analyze numerically the power spectral density of the current
of the ensemble of particles moving in the closed contour and perturbed
by the common for all particles noise. The simplest equations of motion
for such model are of the form
$$
   \dot v = F - \gamma v, \qquad \dot x = v. \eqno(2)
$$
with the perturbation given by the resets of velocity of all particles after
every time interval $\tau$ according to the identical for all particles
replacement ${\bf v}^{new}=\alpha {\bf v}^{old}+{\bf v}_i^{ran}$.
We observe the current power spectral density $S(f)$ dependence on the
frequency $f$ close to the $1/f$--dependence (see Fig.~2).

Our model may be generalized for systems driven by any random forces or
fluctuations. On the other hand, the phenomenon when an ensemble of systems
is linked with a common external noise or fluctuating external fields is
quite usual. Thus, an ensemble of systems in the external random field may
provide a sufficiently universal mechanism of $1/f$--noise.

\noindent {\bf 3. Conclusions}

From the fulfilled analysis we may conclude that, first, synchronization
and transition from chaotic to nonchaotic behavior in ensembles of the
identically perturbed by the random force nonlinear systems may be analyzed
as from the mapping form of equations of motion for the distance between
the particles and the difference of the velocity as well as from the direct
calculations of the Lyapunov exponents and, second, an ensemble of systems
linked with a common external noise may exhibit the $1/f$--type fluctuations.

\noindent {\bf Acknowledgment}

The research described in this publication was made possible in part
by Grant from the Lithuanian State Science and Studies Foundation.

\noindent {\bf References}

\noindent Fahy, S. \& Hamann, D. R. [1992] ``Transition from chaotic
to nonchaotic behavior in randomly driven systems'',
{\it Phys. Rev. Lett.} {\bf 69(5)}, 761-764.

\noindent Kaulakys, B. \& Vektaris, G. [1995a] ``Transition to
nonchaotic behavior in a Brownian-type motion,
{\it Phys. Rev. E} {\bf 52(2)}, 2091-2094.

\noindent Kaulakys, B. \& Vektaris, G. [1995b] ``Transition to nonchaotic
behavior in randomly driven systems: intermittency and 1/f-noise'',
{\it Proc. 13th Int. Conf. Noise in Phys. Syst. and 1/f Fluctuations}
(World Scientific, Singapore), pp.~677-680.

\input{grph.def}
\clearpage

\begin{figure}
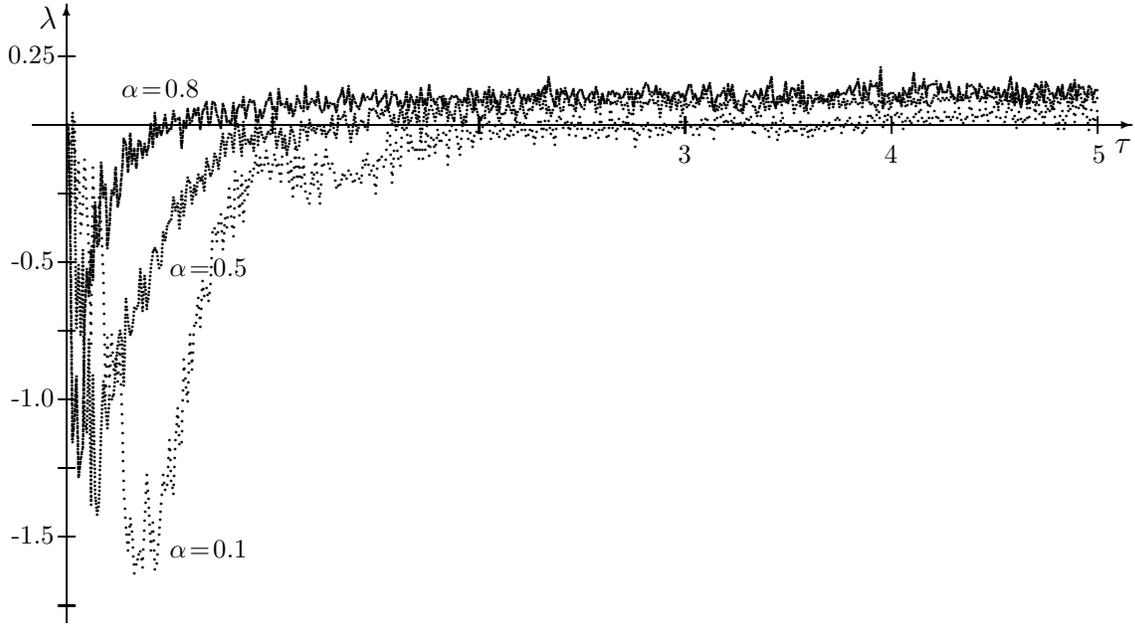

  \begin{picture}(230,184)(-26,-150)
    \input{al01.dat}
    \input{al05.dat}
    \input{al08.dat}
    \put(-10,0){\vector(1,0){320}}
    \put(305,-8){$\tau$}
    \multiput(60,-2.5)(60,0){5}{\line(0,1){5}}
    \put(298.5,-11){\footnotesize 5}
    \put(238,-11){\footnotesize 4}
    \put(178,-11){\footnotesize 3}
    \put(0,-145){\vector(0,1){180}}
    \put(-16,-122){\footnotesize -1.5}
    \put(-16,-82){\footnotesize -1.0}
    \put(-16,-42){\footnotesize -0.5}
    \put(-17,18){\footnotesize 0.25}
    \multiput(-2.5,-140)(0,20){9}{\line(1,0){5}}
    \put(-8,28){$\lambda$}
    \put(30,-126){\footnotesize $\alpha\!=\!0.1$}
    \put(30,-44){\footnotesize $\alpha\!=\!0.5$}
    \put(16,8){\footnotesize $\alpha\!=\!0.8$}
  \end{picture}
  \caption{Lyapunov exponents vs the time $\tau $ between resets of the
  velocity for motion in the Duffing potential according to Eq.~(1)
  with $a=5$, $\omega =1$, $\gamma =0.07$ and different $\alpha$.}
\end{figure}

\begin{figure}
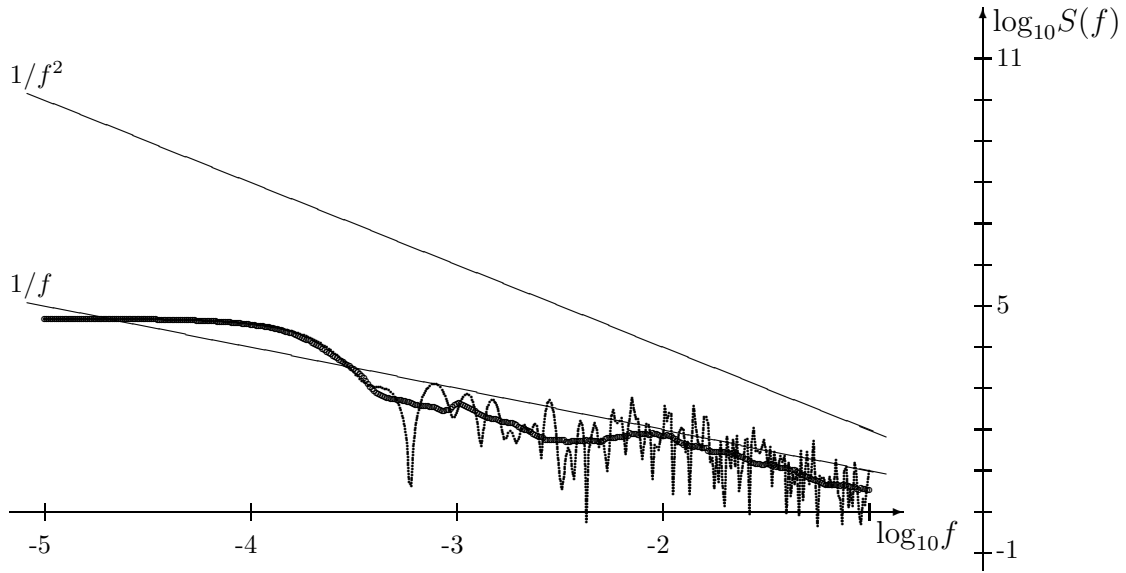

  \begin{picture}(230,184)(-326,-22)
    \input{pert4.dat}
    \def\pl{{\circle{1.5}}}
    \input{pert4av.dat}
    \put(-310,0){\vector(1,0){260}}
    \put(-58,-9){$\log_{10}\!f$}
    \multiput(-300,-2.5)(60,0){5}{\line(0,1){5}}
    \put(-304.5,-12){\footnotesize -5}
    \put(-244.5,-12){\footnotesize -4}
    \put(-184.5,-12){\footnotesize -3}
    \put(-124.5,-12){\footnotesize -2}
    \put(-27,-17){\vector(0,1){164}}
    \put(-23,-13.5){\footnotesize -1}
    \put(-23,58){\footnotesize 5}
    \put(-23,129.8){\footnotesize 11}
    \multiput(-29.5,-12)(0,12){13}{\line(1,0){5}}
    \put(-24,140){$\log_{10}\!S(f)$}
    \put(-305,61){\line(5,-1){250}}
    \put(-305,122){\line(5,-2){250}}
    \put(-310,64){\footnotesize $1/f$}
    \put(-310,125){\footnotesize $1/f^2$}
  \end{picture}
  \caption{The power spectral density of the current of the ensemble of
  particles moving according to Eq.~(2) with $F=1$, $\gamma =0.1$ and
  perturbed by the common for all particles noise
  ${\bf v}^{new}=\alpha{\bf v}^{old}+{\bf v}_i^{ran}$
  with $\alpha =1$ and $\tau =0.1$. The dense line represents
  the averaged spectrum.}
\end{figure}

\end{document}